# Node inspection and analysis thereof in the light of area estimation and curve fitting

A. Kumar
Dept. of Comp. Sc. & Engg.,
Sir Padampat Singhania University
Udaipur, India.
arunkumarsai@gmail.com

P. Chakrabarti
Dept. of Comp. Sc. & Engg.,
Sir Padampat Singhania University
Udaipur, India.
prasun9999@rediffmail.com

P. Saini
Dept. of Comp. Sc. & Engg.,
Sir Padampat Singhania University
Udaipur, India.
poonam.saini9@gmail.com

*Abstract-* **In this paper, we have given an idea of area specification and its corresponding sensing of nodes in a dynamic network. We have applied the concept of Monte Carlo methods in this respect. We have cited certain statistical as well as artificial intelligence based techniques for realizing the position of a node. We have also applied curve fitting concept for node detection and relative verification.**

*Keywords-***Monte Carlo, least square curve fitting.**

## I. INTRODUCTION

As per the coverage area of each node, we divide the total area into 'n' numbers of cells. If the total number of available IP addresses is 'm', then each cell will get m/n number of IP address for communication.

In particular, the first incoming node [1] is treated as a leader and each node has a routing table with a version field. If any node wants to join a particular cell then the leader gives an IP address and updates the routing table as well as incrementing the version field by one and sending the routing table and version field information to the new joining node.

If the leader leaves the cell then the last incoming node acts as a leader. This leads to an investigation towards realizing node within a certain domain. We have to concentrate on extraction of information regarding node coverage in the dynamic network environment.

To understand the complex stochastic systems and to control them, we have to use simulations. Such systems are complex for understanding and need efforts to control, using numerical methods.

Simulations can handle very large complex and realistic systems. In a mobile network, we need to elect a new leader and each connected component of the network has one leader. A mobile adhoc network MANET, sometimes called a mobile mesh network, is a self-configuring network of mobile devices connected by wireless links. Each device is a free to move in a MANET, independently in any direction, therefore its links change with other device frequently.

Each device acts as a router to transfer the data which is not related to its own use. The adhoc network is formed when the nodes are deployed. There is communication amongst the nodes in a network.

To maintain the network, we need to have a leader, so the first and foremost task is to choose a leader. When the group membership changes, there is a change in the group communication protocol as well.

In the system model, let there be a set of 'n' independent mobile nodes. The nodes communicate with each other using the message passing over the wireless network, which is dynamic in nature.

Each node 'i' in the system must have a local variable that holds the identifier of the node currently considered to be the leader of the i's component.

## II. DETERMINATION OF AREA

To estimate the area, Monte Carlo method can be applied. Let us assume that the circle with a unit radius, placed inside a square of side two units. In order to find out whether the node is inside the circle or not in a continuous dynamic environment, we need to calculate the areas of circle and to that of the square.

Area of Circle (area_circle)= $A = \pi r^2$ =$\pi(1)^2$

Area of Square (area_square)=(side)$^2$ =(2)$^2$

In order to find out whether the nodes are inside the circle or not, we need to identify the coordinates P(x,y) where x and y lie in the range [-1,1].

If ($x^2 + y^2$)≤1, then the node is inside the circle. Let **N** be the total number of nodes and **n** be the nodes inside the circle. From Monte-Carlo, we can probabilistically find the number of nodes inside the circular region as:

$$\Rightarrow \frac{area\ of\ unit\ circle}{area\ of\ the\ square} =$$
$$\frac{number\ of\ nodes\ inside\ the\ circle\ (n)}{Total\ number\ of\ nodes\ (N)} \Rightarrow \frac{\pi}{4} = \frac{n}{N}$$

$$\Rightarrow \therefore n = \frac{N \times \pi}{4}$$

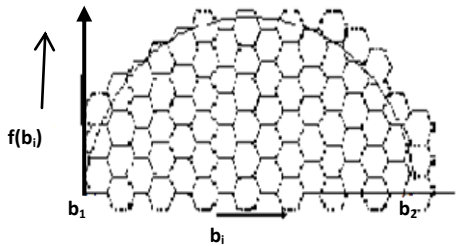

Figure 1. Area Demarcation based on Monte Carlo

We have to generate thousands of random (x,y) positions and determine whether each of them is inside the circle or not. Each time, it is inside the circle the count is incremented, the nodes lying on the circumference of the circle are also part of the final count. Figure 1 shows hexagonal shaped cells in which some of them are partially inside a semicircle, it shows a plot for a range b1 to b2 for a function f(x). After generating a large number of points, the ratio of the number of points inside the circle to the total number of points generated will approach the ratio of the area of the circle to that of the square. The results improve as the value of N is increased maintaining an appropriate proportion. Figure 2 shows simulations done in Matlab by varying N.

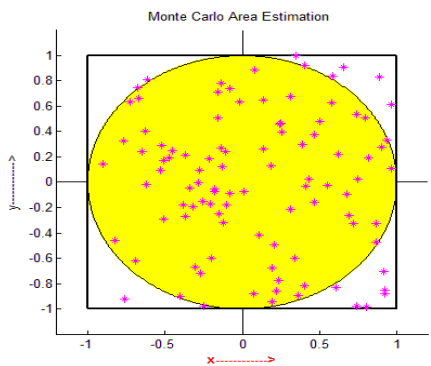

Figure 2. Monte Carlo Area Estimation through Matlab

We assume that in the model total number of available nodes is $N_t$. The nodes with partial portion outside the coverage area will also be treated as nodes of uniform potential as that of those following within the range. Let us assume $N_i$ be the number of incomplete node and $N_c$ be the complete nodes. Therefore, for sensing the area the mathematical expression is given as

$$\frac{(N_i + N_c)}{N_t} = \int_{b_1}^{b_2} f(x)dx\ /\ a(b2 - b1)$$

Figure 3 shows application of Monte carlo method for a rectangle with dimensions 'a' and ($b_2$-$b_1$). For each point, a value of x is selected at random between $b_2$ and $b_1$ say $X_0$. A second random selection is made between 0 and a. If y<=f($X_0$), the point(node) is accepted in the count, otherwise it is rejected, as shown in Fig. 3.

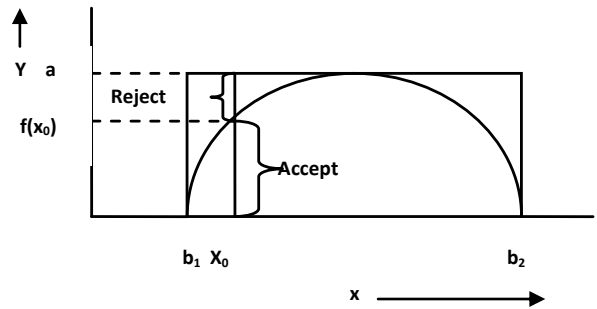

Figure 3: The Monte Carlo Method

If we consider the area for dynamic network as circle then from the center cell, the radius is R for the dynamic network. We calculate the distance for a cell to determine if the cell is in the dynamic network or not, using if $(x_1-x_2)^2/R^2+(y_1-y_2)^2/R^2 = <1$ then the cell is in the radio coverage area.

if$(x_1-x_2)^2/R^2+(y_1-y_2)^2/R^2 =>1$ then the cell is out of the dynamic network, where $(x_1,y_1)$ is the center point of the center node and $(x_2,y_2)$ is the center point for any node.

### III. NODE PREDICTION IN THE LIGHT OF GRAPHICAL PLOTTING

If the matched information elements are arranged in random order or haphazardly then there is some method for prediction of the position of each element based on some curvical nature. We assume that the elements follow the equation of some normal curve. Hence the position can be calculated by changing the variables of the equation of the curve at different timing instant. We can even do clustering, so that, instead of element detection we can concentrate on cluster detection, thereby minimizing search overhead.

We can predict the position of a particular information element if there is a relation between the position of a particular timing instant and the value of the timing instant [2]. If we have the idea of position of element at two timing intervals then we can predict value of the position of the element both at the midway position and also the extreme position, provided the matched elements are in equidistant form. If they are not in equidistant form, in that case the incoming element is examined with respect to the curve fitting between the earlier elements in accordance with exponential curve form. N's represent the information elements while cv's represent the curves.

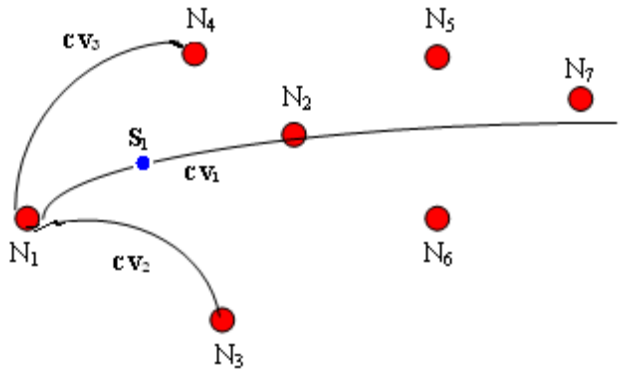

Figure 4. Curvical nature in the light of exponential form.

**Theorem 1**: *If there bears a relation between positions of a element with a particular timing instant, then the same relation is valid, if the element is present in the midway position.*

*Proof:* Let the equation for determination of element is $P_{N,t} = Me^{at}$ where $P_{N,t}$ is position of element at a time instant 't'.

'M' and 'a' are constants.

We further assume that elements are $N_1$, $N_2$, $N_3$, …. and so on.

Therefore, $P_{N1,t1} = Me^{at1}$ and $P_{N2,t2} = Me^{at2}$

So, the midway interval time is $(t_1+t_2)/2$.

Now, $(P_{N1,t1} \cdot P_{N2,t2})^{1/2} = [(Me^{at1}) \cdot (Me^{at2})]^{1/2}$
$$= [M^2 e^{a(t1+t2)}]^{1/2}$$
$$= Me^{a(t1+t2)/2} = P_{S1,(t1+t2)/2}$$

Hence, we conclude that the same relation is valid in midway position, in this case $S_1$ indicates the midway position of the curve connecting $N_1$ and $N_2$ elements.

**Theorem 2**: *To determine the position of an element in respect of first curve, concept of exponential growth or decay is valid.*

*Proof:* Let the first curve $CV_1$ passing through elements $N_1$ and $N_2$ is exponential curve satisfying equation,

$$P_{N,t} = Me^{at}.$$

Now element $N_3$ has to be examined with respect to curve $CV_1$, whether it is exponential growth (charging curve) or decay (discharging curve) curve emanating from node $N_1$.

In Figure 1, the node $N_1$ is connected to element $N_3$ and the curve is exponential decay type, satisfying the equation, $P_{N3,t3} = Me^{-at3}$.

Therefore the curve fitting between the elements $N_1$ and $N_3$ is $CV_2$ exponential decay curve consisting of new element $N_3$.

Similarly, in case of element $N_4$, it is exponential growth type, satisfying $P_{N4,t4} = Me^{at4}$ equation.

**Theorem 3:** *If a particular timing instant the position of an element is in form of arithmetic mean of two intervals and position of another element is in the form of their harmonic mean, then square root of the product of these two means reveals the geometric mean of the two timing instants.*

*Proof:* We assume that $t_1$ and $t_2$ are two timing instants. Let at t' arithmetic mean is valid act at t", harmonic mean is valid.

$P_{N,t}' = (t_1+t_2)/2$ and $P_{N,t}" = 2/\{(1/t_1)+(1/t_2)\}$

Hence, $P_{N,t}' \cdot P_{N,t}" = t_1 \cdot t_2$

$(P_{N,t}' \cdot P_{N,t}")^{(1/2)} = (t_1 t_2)^{(1/2)}$, which is geometric mean of $t_1$ and $t_2$.

## IV. NODE REALIZATION ON THE BASIS OF LEAST SQUARE FITTING

### A. Concept

A mathematical procedure for finding the best fitting curve to a given set of points is by minimizing the sum of the squares of the offsets of the points from the curve. There can be vertical offsets and perpendicular offsets.

The linear least square fitting technique provides a solution of the best fitting straight line through a set of node points (x, y) indicating their physical presence in a network.

In practice, the vertical offsets from a line are almost minimized instead of the perpendicular offsets. This provides a fitting function for the independent variable **t** that estimates **y** (position) for a given value of **t**.

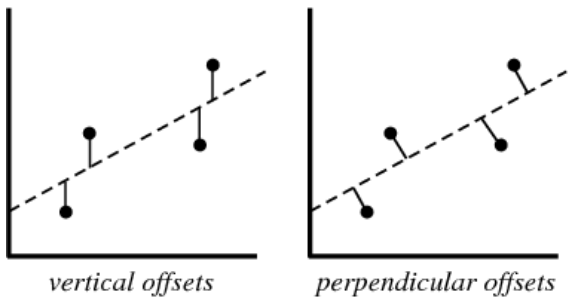

Figure 5. Linear Fitting Curve depicting nodes for a dynamic network with vertical and perpendicular offsets.

The residuals of the best fit line for a set of **n** nodes using unsquared perpendicular distance $d_i$ is given by:

Case 1: Perpendicular Offsets

$$R_\perp \equiv \sum_{i=1}^{n} d_i$$

Since, the perpendicular distance from a line **y = a + bx** to a point **i** is given by

$$d_i = \frac{|y_i - (a + bx_i)|}{\sqrt{1 + b^2}}$$

We can calculate the distance of points (nodes) **i** with coordinate (x,y) from a line with equation **y = a + bx.** Here 'b' =slope and 'a' is the perpendicular offset of the point (node) from a line [3][4][5].
The function to be minimized is

$$R_\perp = \sum_{i=1}^{n} \frac{|y_i - (a + bx_i)|}{\sqrt{1 + b^2}}$$

Square of the perpendicular distance, we get

$$R_\perp^2 = \sum_{i=1}^{n} \frac{[y_i - (a + bx_i)]^2}{(1 + b^2)}$$

This can be minimized as

$$\frac{dR_\perp^2}{da} = \frac{2}{1+b^2} \sum_{i=1}^{n} (y_i - (a + bx_i))(-1) = 0 \quad \text{-----(1)}$$

$$\frac{dR_\perp^2}{db} = \frac{2}{1+b^2} \sum_{i=1}^{n} (y_i - (a+bx_i))(-x_i) + \sum_{i=1}^{n} \frac{[y_i - (a+bx_i)]^2(-1)(2b)}{(1+b^2)^2} = 0$$

For a line **y=ax+b**, we have here 'a' which is the perpendicular offset value.

$$a = \frac{\sum_{i=1}^{n} y_i - b \sum_{i=1}^{n} x_i}{n}$$

From the eq. (1), we get

$$\Rightarrow \overline{y} - b\overline{x} \text{ and 'b'(slope) is given by:}$$

$$b = -B \pm \sqrt{B^2 + 1}$$

where

$$B = \frac{\frac{1}{2}\left(\sum_{i=1}^{n} y_i^2 - n\overline{y}^2\right) - \left(\sum_{i=1}^{n} x_i^2 - n\overline{x}^2\right)}{n\overline{x}\overline{y} - \sum_{i=1}^{n} x_i y_i}$$

Case2: Vertical offset

In the case of vertical least square fitting function for the sum of the squares of the vertical deviation $R^2$ of a set of n nodes

$$R^2 = \sum_{i=1}^{n} [y_i - f(x_i, a_1 a_2 .... a_n)]^2$$

$$\frac{dR^2}{da_i} = 0 \text{ For i=1.......n for a linear fit}$$

$f(a,b) = a + bx$ where a is the intercept and b is the slope.

$$\Rightarrow R^2(a,b) \equiv \sum_{i=1}^{n} [y_i - (a + bx_i)^2]$$

$$\Rightarrow \frac{dR^2}{da} = -2\sum_{i=1}^{n}[y_i - (a+bx_i)] = 0$$

$$\Rightarrow \frac{dR^2}{db} = -2\sum_{i=1}^{n}[y_i - (a+bx_i)]x_i = 0$$

From the above we get,

$$\Rightarrow na + b\sum_{i=1}^{n}x_i = \sum_{i=1}^{n}y_i$$

$$\Rightarrow a\sum_{i=1}^{n}x_i + b\sum_{i=1}^{n}x_i^2 = \sum_{i=1}^{n}x_i y_i \quad \{\text{multiplying by } x_i, \text{both sides}\}$$

The matrix representation is as follows:

$$\begin{bmatrix} n & \sum_{i=1}^{n}x_i \\ \sum_{i=1}^{n}x_i & \sum_{i=1}^{n}x_i^2 \end{bmatrix} \begin{bmatrix} a \\ b \end{bmatrix} = \begin{bmatrix} \sum_{i=1}^{n}y_i \\ \sum_{i=1}^{n}x_i y_i \end{bmatrix}$$

$$\begin{bmatrix} a \\ b \end{bmatrix} = \begin{bmatrix} n & \sum_{i=1}^{n}x_i \\ \sum_{i=1}^{n}x_i & \sum_{i=1}^{n}x_i^2 \end{bmatrix}^{-1} \begin{bmatrix} \sum_{i=1}^{n}y_i \\ \sum_{i=1}^{n}x_i y_i \end{bmatrix}$$

We find out the sum of the squares i.e. SSxx, SSyy, SSxy values.
The sum of the squares is given by:

$$\therefore SSxx = \sum_{i=1}^{n}(x - \bar{x})^2$$

$$= \left(\sum_{i=1}^{n}x_i^2\right) - x\bar{x}^2$$

$$SSyy = \sum_{i=1}^{n}(y_i - \bar{y})^2$$

$$= \left(\sum_{i=1}^{n}y_i^2\right) - x\bar{y}^2$$

$$SSxy = \left(\sum_{i=1}^{n}x_i y_i\right) - n(\bar{x})(\bar{y})$$

σ is the standard deviation, it is the positive square root of the arithmetic mean of the square of the distance of the given values from their arithmetic mean.

$$\sigma = (\frac{1}{N}\Sigma_i fi((xi - \overline{x}))^2)^{1/2}$$

$\bar{x}$ is the arithmetic mean of the distribution and $\Sigma_i fi = N$. The square of the standard deviation is called variance and is given by:

$$\sigma^2 = \frac{1}{N}\Sigma_i fi((xi - \overline{x}))^2$$

$$\sigma_x^2 = \frac{SSxx}{n}$$

$$\sigma_y^2 = \frac{SSyy}{n}$$

$$Cov(x, y) = \frac{SSxy}{n}$$

b is the regression coefficient, regression analysis is the average relationship between two or more variable in terms of the original units of the data. The line of regression is the line of the best fit.

$$b = \frac{Cov(x, y)}{\sigma_x^2} = \frac{SSxy}{SSxx}$$

$$a = \bar{y} - b\bar{x}$$

Overall quality fit $\quad r^2 = \frac{SS^2 xy}{SSxx SSyy}$

We plot the curve for the sample data in Matlab and the same has been shown in Figure 5, Figure 6, Figure 7. The standard error in obtaining the values of a vertical offset of a point (node) in a dynamic environment is given as under:

$$SE(a) = s\sqrt{\frac{1}{n} + \frac{\bar{x}^2}{SSxx}}$$

The error in the value of 'b' i.e. the slope of the line is given as below.

$$SE(b) = \frac{s}{\sqrt{SSxx}}$$

*B. Sensing analysis*

For a best fit line the value of r ("goodness of fit" of the least squares line, called the coefficient of correlation) should be approximately equal to 1 or -1, for a worst fit the value approaches towards zero.

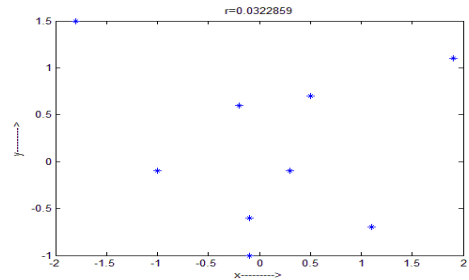

Figure 6. Scattered points for r=0.0322859

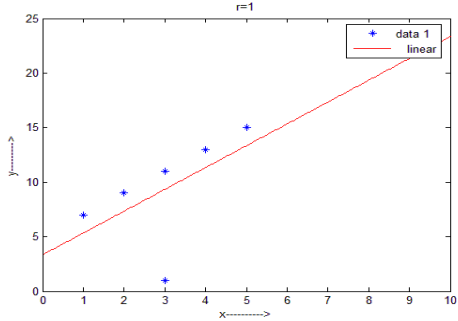

Figure 7. Best fit line for r=1

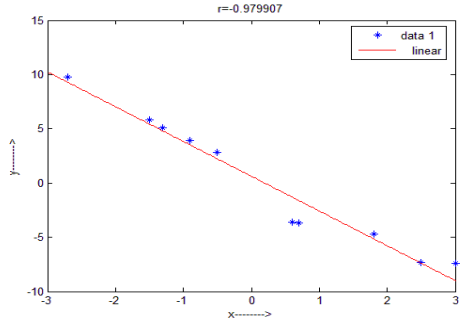

Figure 8. Best fit line with r= - 0.979907

Correlation coefficient indicates the strength and direction of a *linear* relationship between two random variables. In general statistical usage, *correlation* or co-relation refers to the departure of two random variables from independence.

The closer the coefficient is to either −1 or 1, the stronger the correlation between the variables. If the variables are independent then the correlation is 0, but the converse is not true, because the correlation coefficient detects only linear dependencies between two variables. For a best fit line for r=1, we see in Figure 7. that the nodes (points) are best fitted over a linear curve. Similarly in Figure 8, for r= - 0.979907, the nodes are best fitted. But in the case of r=0.0322859, nodes are randomly distributed as shown in Figure 6.

V. EXPONENTIAL BEHAVIOR OF POSITIONAL NODES AND ANALYSIS

*A. Concept*

The three basic exponential models to asses any system where the number of entries increase and decrease exponentially in a system include:
(i)Exponential Growth model
(ii)Exponential Decay Model
(iii)Modified Exponential Growth Model

(i)Exponential Growth Model

The word growth refers to a rate of change and growth involves differential equations. The rate of growth of the number of nodes in a dynamic network is increasing at the rate of coefficient 'k' of the present population.

$\dot{y} = k*y$, $y = y_0$ at time t = 0

This is a first order differential equation whose solution is the exponential function.

$$\therefore y = y_0 e^{kt}$$

The curve for the exponential growth shows different values of k and $x_0=1$. The curve clearly depicts that the fact that the growth rate of the nodes in a dynamic environment is directly proportional to the current level, as shown in Figure 9.

(ii)Exponential Decay Model
Another way of describing the exponential growth is the exponential decay where the variable decays from the initial value $y_0$ at a rate proportional to the current rate. The model can also be called as negative growth model and the equations are:

$\dot{y} = - k*y$

$y = y_0$ at time t = 0, the solution of the exponential function is

$$\therefore y = y_0 e^{-kt}$$

The decay model is applicable to dynamic networks as the number of nodes in a region decrease with time. Therefore there is decay in the system, as shown in Figure 10.

(iii)Modified Exponential Growth Model:

The number of nodes in a dynamic environment as per the exponential growth model can increase up to any limits. But actually it is not so, as there cannot be an unlimited growth. For a dynamic environment, the growth rate is proportional to the number of nodes which have not yet entered the service region. Let 'N' be the number of nodes which might enter the service region, and let 'n' be the nodes which are already there in the service region. Then the number of nodes which have yet not entered shall be (N-n) and let k be the coefficient of proportionality, then the modified exponential growth model can be represented with equations as:

$$\dot{n} = k(N-n)$$

And when t=0, n=0; and the solution is:

$$n = N(1 - e^{-kt}).$$

The curve shows that that the origin maximum slope occurs and slope decreases as the time increases. Therefore the curve approaches the limit more slowly, but it never reaches the limit. Therefore we can conclude that the number of nodes entering the service region drops as the time passes, as shown in Figure 11.

*B Graphical analysis*

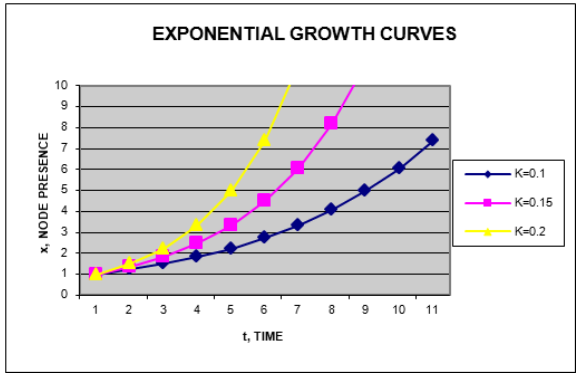

Figure 9. Exponential growth Curve (for k=0.1,0.15, 0.2)

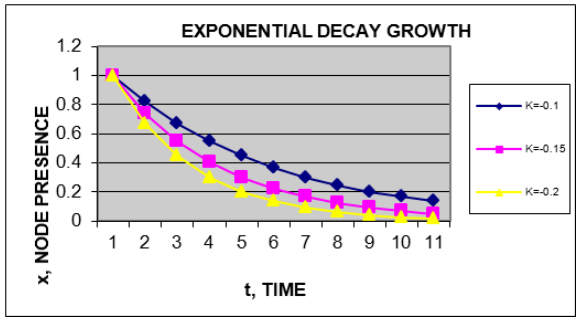

Figure 10. Exponential Decay Growth(for k=-0.1, -0.15,-0.2)

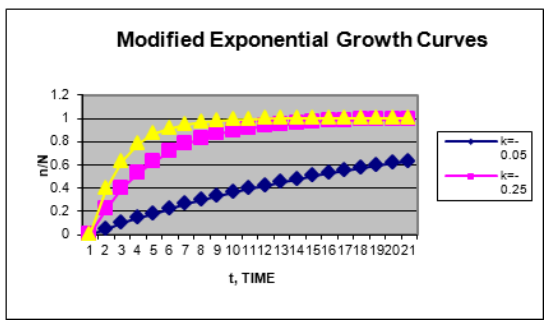

Figure 11. Modified Exponential Growth Curve(for k=0.05,0.25,0.5)

VI. CONCLUSION

Identification of a node in the light of predictive analysis plays a pivotal role in dynamic network model. Monte Carlo technique has been applied for effective cell detection. We have also analyzed position of nodes using statistical analysis based on the concept of curve-fitting. Respective analysis for vertical as well as perpendicular offset in terms of least square fitting has also been depicted in this paper. The paper also points out, the exponential behavior of the positional nodes and the corresponding graphical representation has also been embodied herewith.